\documentstyle[twocolumn,prc,aps]{revtex}
\input epsf
\begin{document}
\twocolumn[\hsize\textwidth\columnwidth\hsize\csname @twocolumnfalse\endcsname

\title{Baryon Fluctuations and the QCD Phase Transition}

\author{David Bower and Sean Gavin}
\address{
Department of Physics and Astronomy, Wayne State University, 
Detroit, MI, 48202}
\date{\today} 
\maketitle
\begin{abstract}
  The dynamic separation into phases of high and low baryon density in
  a heavy ion collision can enhance fluctuations of the net rapidity
  density of baryons compared to model expectations.  We demonstrate
  that event-by-event proton and antiproton measurements can be used
  to observe this phenomenon.  We then perform real-time lattice
  simulations to show how these fluctuations arise and how they can
  survive through freeze out.

\vspace{0.1in}
\end{abstract}
]

\begin{narrowtext}

If the QCD phase transition is first order, matter at the appropriate
temperatures and densities can form a mixed phase consisting of plasma
droplets in equilibrium with a surrounding hadronic fluid.  If formed
in ion collisions, this mixed phase can produce large event-by-event
fluctuations as the system hadronizes \cite{Mishustin}. In particular,
extraordinary baryon number fluctuations \cite{Gavin} can accompany
the first order transition expected at high baryon density
\cite{Rajagopal}.

In this paper we explore the dynamics of phase separation in nuclear
collisions.  The aims of this paper are twofold. First, we study the
role of baryon number fluctuations as a probe of the order of the QCD
transition. We focus on the high baryon density regime, where theory
\cite{Rajagopal} and lattice simulations \cite{Fodor} suggests that
the QCD phase transition is first order in a strict thermodynamic
sense with baryon density as an order parameter. Our work may also
apply to RHIC collisions, if the low baryon density systems produced at
the highest energies {\it approximate} a first order transition
\cite{Gavai,McLerran}.  Second, we generalize techniques from
condensed matter physics \cite{Bray} to confront phase separation in the
highly-nonequilibrium context of nuclear collisions. Our framework can
be used to systematically address other probes as experimental
information and theoretical understanding evolve. 

To begin, we describe the character of mixed-phase baryon fluctuations
and show how they can be measured. Measurement is not completely
straightforward as, e.g., neutrons are not easily observed on an
event-by-event basis. We then formulate a dissipative-hydrodynamic
model of phase separation and perform numerical simulations for that
model.
  
QCD with two massless flavors can exhibit a first order transition
whose coexistence curve culminates in a tricritical point at
temperature $T_c$ and baryon chemical potential $\mu_c$
\cite{Rajagopal}.  For $T>T_c$ and $\mu<\mu_c$, a second order phase
transition breaks/restores chiral symmetry. If the quark masses are
sufficiently large, the second order transition is replaced by a
smooth transformation (since chiral symmetry is explicitly
broken). The first order line remains, however, with the tricritical
point replaced by a critical point in the same universality class as a
liquid--gas transition. 
  
At RHIC, baryon density may also serve as an approximate order
parameter for the nearly first order transition at small net baryon
density. Lattice simulations \cite{Bernard,Gavai} and general
arguments \cite{McLerran,Bernard,Kunihiro} show that the baryon
susceptibility $\chi$ at $\mu = 0$ can increase suddenly as
temperature is increased near $T_h\sim 160$~MeV, where the chiral
order parameter and, e.g., the energy density change sharply. Jumps in
the susceptibility commonly accompany first order transitions. For a
liquid-gas transition, $\chi = \partial\rho/\partial \mu$ is
proportional to the compressibility: steam is much more compressible
than water.

Large fluctuations in baryon number occur during phase separation in a
first order transition. Figure 1b shows the phase diagram in the
$T-\rho$ plane \cite{Rajagopal}, where $\rho$ is the baryon density. A
uniform system quenched into the outer parabolic region will separate
into droplets at the high baryon density $\rho_q$ surrounded by matter
at density $\rho_h$. The net baryon number $N_B$ in a sub-volume of
the system varies depending on the number of droplets in the
sub-volume. The variance of the baryon number $V_B= \langle
N_B^2\rangle - \langle N_B\rangle^2$ can exceed the equilibrium
expectation by an amount
\begin{equation}\Delta V_B \approx f(1-f)(\Delta N_B)^2,
\label{eq:fluct}\end{equation}
where $f$ is the fraction of the high density phase in the sub-volume
$V$ and $\Delta N_B = (\rho_q - \rho_h)V$.  In contrast, an
equilibrium system follows Poisson statistics, so that $V_B = V +
\overline{V} = \langle N+{\overline N}\rangle$, where $N$, $V$ and
$\overline N$, $\overline V$ are the numbers and variances of baryons
and antibaryons and $N_B = N - \overline{N}$.

Experimenters can search for a ``super-poissonian'' variance such as 
(\ref{eq:fluct}) by measuring
\begin{equation}
\Omega_{p} = 
{{V_{p-\overline{p}}-\langle N_p + N_{\overline p}\rangle}
\over
{\langle N_p +  N_{\overline p}\rangle^2}},
\label{eq:Omega}\end{equation}
where $N_p$ and $N_{\overline p}$ are the numbers of protons and
antiprotons in a rapidity interval and $V_{p-\overline{p}}$ is the
variance of the net proton number $N_p - N_{\overline p}$.  This
quantity vanishes in equilibrium and is related to the more familiar
scaled variance $\omega_p =
\langle N_p + N_{\overline p}\rangle(1+\Omega_{p})$.
Most importantly, $\Omega_{p}$ is ideal
for our application because of the property
\begin{equation}
\Omega_{p} = \Omega_B \equiv 
{{V_{B}-\langle N + \overline{N}\rangle}
\over
{\langle N + \overline{N} \rangle^2}} 
\label{eq:Omega0}\end{equation}
where $N$ and $\overline N$ are the numbers of baryons and antibaryons
-- including unseen neutrons and antineutrons (the proof follows). The
conditions for which (\ref{eq:Omega0}) holds are met by a range of
thermal and Glauber models that respect isospin symmetry. Isospin
fluctuations can alter (\ref{eq:Omega0}) near the tricritical point or
in the presence of a disoriented chiral condensate, but those effects
will be evident from pion measurements.

We demonstrate (\ref{eq:Omega0}) by writing the joint probability for
$N_p$ and $N_{\overline{p}}$ as $\sum_{N,{\overline N}} p(N_p|N)
p(N_{\overline p}|\overline{N})P(N,{\overline N})$. The distribution
$P(N,{\overline N})$, which determines $\Omega_B$, is modified by
phase separation; we make no assumptions about its form. We assume
that the conditional probability $p(N_p|N)$ for measuring $N_p$ given
$N$ baryons is binomial, with $q$ the chance that any individual
baryon is a proton (see \cite{Cowen} for notation). We further take
$p(N_{\overline p}|{\overline N})$ for antiprotons to be binomial with
the same $q$. These assumptions hold for most thermal and
multiple-scattering models.  The average of the joint distribution is
$\langle N_p + N_{\overline{p}}\rangle = \sum_{N,{\overline N}}
p(N,{\overline N}) (\nu+\overline{\nu})$, where the binomial averages
$\nu=\sum_{N_p}p(N_p|N)N_p$ and $\overline{\nu}=
\sum_{N_{\overline {p}}}p(N_{\overline{p}}|\overline{N})N_{\overline{p}}$ 
yield $\langle N_p + N_{\overline{p}}\rangle = q\langle N +
\overline{N}\rangle$. The quantity $\langle N + \overline{N}\rangle$
depends only on $P(N,\overline{N})$. Similarly, we find $\langle (N_p -
N_{\overline{p}})^2\rangle = q^2\langle (N-\overline{N})^2\rangle +
q(1-q)\langle N + \overline{N}\rangle$.
We combine these moments to obtain (\ref{eq:Omega0}).

The antiproton contribution to (\ref{eq:Omega}, \ref{eq:Omega0}) is
large only at RHIC, where $N_{\overline p}/N_p
\sim 0.6$ at $\sqrt{s}=130\, A\cdot$GeV \cite{RHICpbar}.  At the top
SPS energy, we estimate $\overline p$ contributions to
(\ref{eq:Omega}) to be at the few percent level in Au+Au at
$\sqrt{s}=17.5\, A\cdot$GeV, since $N_{\overline p}/N_p \sim 6\%$
\cite{NuXu}. The highest baryon density -- and the greatest potential
for observing a first order transition -- is perhaps at lower
energies.

We remark that Jeon and Koch and Asakawa {\it et al}. have proposed that
hadronization may change the character of charge and baryon number
fluctuations even in the absence of a phase transition
\cite{Koch}. This effect is essentially poissonian, however,
so it is not clear that it would cause $\Omega_{p}$ to
differ from zero, the equilibrium value, or that it could be tested
without measuring neutrons. The effect on charge fluctuations is much
more dramatic
\cite{Koch}.

We now turn to describe the process of phase separation. To describe
the state of the mixed phase, we follow the standard condensed matter
practice \cite{Bray} and write a Ginzburg-Landau free energy
$f=\kappa(\nabla \rho)^2/2 + f_0$, where
\begin{equation}
f_0 = -m^2(\rho-\rho_c)^2/2 + \lambda (\rho-\rho_c)^4/4
\label{eq:phi4}\end{equation}
describes the excursions of the baryon density
$\rho$ from its equilibrium value in the uniform matter. For
$m^2\propto T_c-T$ we find the correct liquid-gas critical
exponents. The values $\rho_h$ and $\rho_q$ in fig. 1
correspond to the equilibrium densities at $T < T_c$: $\rho_h =
\rho_c - \Delta\rho$ and $\rho_q = \rho_c + \Delta\rho$, where $\Delta\rho 
= \sqrt{m^2/\lambda}$. The $\kappa$ term describes the droplet surface
tension. For our $f_0$, we compute $\sigma = (8\kappa
m^3/9\lambda^2)^{1/2}\propto\kappa^{1/2}$ \cite{Widom}.

To describe the dynamics of the system, we must account for the fact
that baryon number is conserved. Furthermore, it is crucial to include
dissipation to describe this strongly fluctuating system. The simplest
equations that meet these criteria are:
\begin{equation}
\partial \rho/\partial t = M\nabla^2\mu, \,\,\,\,\,\, \,\,\,\,\,\, 
\mu = f_0^\prime -\kappa\nabla^2\rho;
\label{eq:diff}
\end{equation}
model B in \cite{Bray}. We illustrate that (\ref{eq:diff}) describes 
diffusion in a stable liquid by considering fluctuations about the 
equilibrium density $\rho = \rho_h +
\delta\rho_k \exp(-i{\bf k}\cdot{\bf x})$, where 
$\delta\rho_k \ll \rho_h$. A system at this density is near the minimum
of $f_0$, so that $f_0^\prime \approx
f_0^{\prime\prime}(\rho_h)\delta\rho_k = 2m^2 \delta\rho_k$. Therefore,
(\ref{eq:diff}) is standard diffusion equation at linear order in
$\delta\rho_k$.  We identify the baryon diffusion coefficient at
$\rho_h$ as $D=2m^2M$. In general, diffusion drives the system
towards homogeneity at all density for which
$f_0^{\prime\prime}(\rho)>0$.
\begin{figure} 
\epsfxsize=3.25in
\leftline{\epsffile{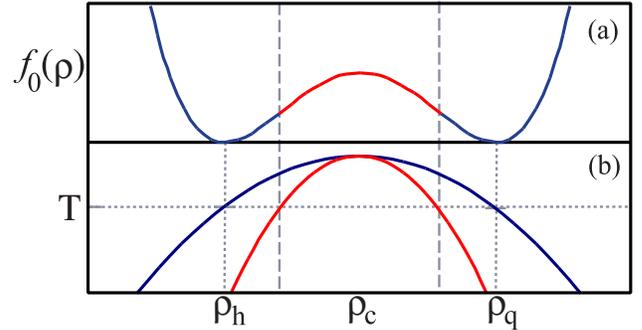}}
\caption[]{
  Free energy (a) and phase diagram (b) vs. baryon density for
  (\ref{eq:phi4}).}\end{figure}

Phase separation is most dramatic if the rapid expansion of the heavy
ion system drives the system into the unstable region; i.e., the inner
parabolic region in fig.~1b, corresponding to
$f_0^{\prime\prime}(\rho) < 0$ in fig.~1a. Droplets form from runaway density
fluctuations in a process known as spinodal decomposition.  We
estimate the time scale $\tau_R$ for this process by considering the
time evolution of small fluctuations near $\rho_c$. We take
$\rho = \rho_c +
\delta\rho_k \exp(-i{\bf k}\cdot{\bf x})$ to find
\begin{equation}
{{d}\over{d\tau}}\delta\rho_k = 
m^2M\left(k^2 - {{\kappa}\over{m^2}}k^4\right)\delta\rho_k
\equiv {{\delta\rho_k}\over {\tau_k}}
\label{eq:linear}\end{equation}
to linear order in $\delta\rho$. Long wavelength disturbances
corresponding to $0< k < \sqrt(m^2/\kappa)$ grow with time, while the
surface-tension term stabilizes the shorter wavelength modes. The time
scale for growth $\tau_k$ is shortest at $k_R = \sqrt(m^2/2\kappa)$. 

The fastest-growing mode at wavenumber $k_R$ dominates the early
evolution of the system in the unstable regime.  The time scale for
the growth of this mode is
\begin{equation}
\tau_R = 8\xi^2/D
\label{eq:time}\end{equation}
where $\xi =\kappa^{1/2}/m$ is the correlation length. 
The values of
$\xi$ and $\tau_R$ determine the time and spatial scales for the onset
of spinodal decomposition. We plausibly estimate the correlation
length to be $\xi\sim 1$~fm, roughly the value of the inverse sigma
mass. For a value of $D\sim 8$~fm consistent with calculations in
\cite{Prakash}, we find $\tau_R\sim 1$~fm as \cite{Gavin}. We remark that
the large magnitude of $D$ suggested by \cite{Prakash} is consistent
with our assumption in (\ref{eq:diff}) that baryon diffusion is the
dominant transport mode for baryons at high density. Our model
superficially suggests a slower onset of the instability for a
substantially smaller value of $D$. However, if $D$ were truly small
then it would be necessary to include transport mechanisms involving
convection and viscosity. In fact, viscosity must dominate near $\mu =
0$, where diffusion precesses are strictly irrelevant \cite{Gavin2}.

To describe nuclear collisions, we extend (\ref{eq:diff}) to include
drift due to Bjorken longitudinal flow:
\begin{equation}
\partial \rho/\partial \tau + \rho/\tau = M\nabla^2\mu,
\label{eq:drift}\end{equation}
where $\tau$ is the proper time and $\mu$ is given by (\ref{eq:phi4},
\ref{eq:diff}). The new drift term forces the average density to decrease as
$\langle\rho\rangle\propto \tau^{-1}$, driving the system through the
phase coexistence region. Fluctuations grow when densities approach 
$\rho_c$ (see fig.~1). To derive the drift term, observe that
(\ref{eq:diff}) follows from baryon current conservation, which
more generally implies $\partial_\mu j^\mu =0$.  The current is $j^\mu
= \rho u^\mu + j_d^{\mu}$ \cite{LL}, where $u^{\mu}$ is a fluid
velocity that includes a contribution from the meson flow, and $j_d$
is the diffusion current, $\propto \nabla\mu$ when $u =
(1,0,0,0)$. The left and right sides of (\ref{eq:drift}) respectively
follow from $\partial_\mu (\rho u^\mu)$ and $\partial_\mu j_d^\mu$ for
Bjorken flow.

For times $t\gg \tau_R$, the system undergoes a nonlinear evolution in
which droplets merge, reducing their surface energy. To study this
regime, we write the evolution equation (\ref{eq:drift}) in
the dimensionless form:
\begin{equation}
{{\partial\psi}\over{\partial \hat{\tau}}} + {{\epsilon + \psi}\over
{\hat{\tau}}} = -{{1}\over{2}}\hat{\nabla}^2(\psi -\psi^3 +
\hat{\nabla}^2\psi)
\label{eq:dim}\end{equation}
where we use the dimensionless coordinates $\hat{\tau} = 8\tau/\tau_R$
and $\hat{\bf x} = {\bf x}/\xi$. The dimensionless order
parameter $\psi \equiv (\rho - \rho_c)/\Delta\rho$ equals $\pm 1$
when $\rho = \rho_{h,\, q}= \rho_c
\pm \Delta\rho$. The only remaining parameter is $\epsilon =
\rho_c/\Delta\rho$, which controls the strength of the first order
transition. Here, we take $\epsilon = 1$ corresponding to a strongly
first order transition. Observe that (\ref{eq:dim}) depends on the
temperature and density scale only through $\epsilon$. This is an
artifact of our very simplistic quadratic $f_0$; we will introduce a
more realistic free energy density in later work to study the role of
temperature in the evolution.

We solve (\ref{eq:dim}) numerically on a 2+1 dimensional lattice
following Grant {\it et al.} \cite{Grant}. We use a forward Euler
method to evolve the system in time for a time step $\Delta\hat{\tau} =
0.05$. We study the evolution in the transverse plane and in the rapidity
$\eta$-$x_T$ plane, where $x_T$ is a cartesian transverse coordinate. The 
laplacian in the $\eta$-$x_T$ case is  
\begin{equation}
\hat{\nabla}^2 = {{1}\over{\hat{\tau}^2}}{{\partial^2}\over{\partial \eta^2}}
+ \hat{\nabla}_\perp^2.
\end{equation}
To treat the higher spatial derivatives we extend the
next-nearest-neighbor algorithm developed by Oono and Puri
\cite{Puri} and used in \cite{Grant} to account for the asymmetric
$\eta$-$x_T$ lattice. We write
\begin{equation}
\hat{\nabla}^2\psi = {{1}\over{2(\Delta x)^2}}
\left(\sum_{NN} \psi + 
{{1}\over{4}}\sum_{NNNN} \psi 
- {{5}\over{2}} \psi\right)
\label{eq:oono}\end{equation}
where the first sum runs over the four nearest neighbors (NN) and the
second over the four adjacent next-next-nearest neighbors (NNNN). Oono
and Puri use the diagonal next-nearest-neighbors instead in
(\ref{eq:oono}) -- a formulation that requires a symmetric lattice.  We
take $\Delta x = 1$. We find that our results are practically
indistinguishable from NNN results \cite{Grant} for this spacing
on a symmetric lattice. To study longitudinal expansion, it suffices
to replace one coordinate $\Delta x$ for $\hat{\tau}\Delta\eta$.

\begin{figure} 
\epsfxsize=3.0in
\centerline{\epsffile{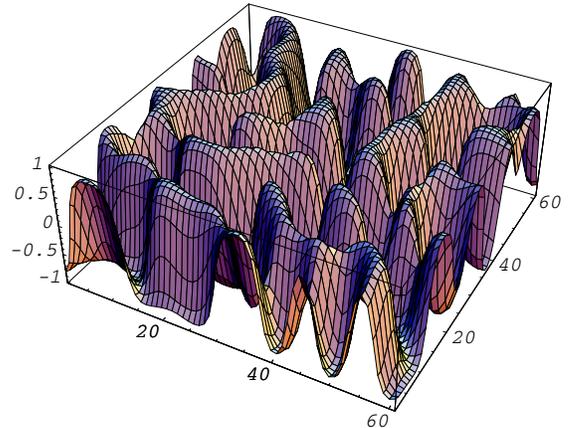}}
\caption[]{
  Order parameter in the transverse plane in the absence of expansion.
Droplets tend to merge.
}\end{figure}
Figures 2 and 3 show 2+1 dimensional numerical simulations of
(\ref{eq:dim}) in the transverse plane. Only longitudinal expansion is
considered so the coordinates are cartesian with periodic boundary
conditions.  For comparison, fig.~2 shows results in which expansion
is neglected by omitting the term $(\epsilon+\psi)/\hat{\tau}$ in
(\ref{eq:dim}). Expansion shown in fig.~3 prevents droplets from
merging as in fig.~2. The expanding system reaches $\rho_c$ at
$\tau_0 = 5$~fm. Because this is a dissipative system, we must apply
thermal noise at each lattice site at $\tau_0$ to seed phase
separation (noise at earlier times is dampened). The memory of the
initial conditions is essentially lost for $\tau - \tau_0 > \tau_R$.

We now study the rapidity dependence of baryon number
fluctuations. Figure 3 shows the computed variance for two different
initial times and for two rapidity intervals. The variance is computed
from a sample of 5000 simulated events, each unique due to the thermal
noise. We see that the super-poissonian fluctuations grow appreciably
by $\tau
\sim 2\tau_0$. This variance drops as the rapidity interval is
increased. We find that variance is governed by the ratio
$\tau_0/\tau_R$, which compares the expansion and droplet-growth time
scales. 
\begin{figure} 
\epsfxsize=3.0in
\centerline{\epsffile{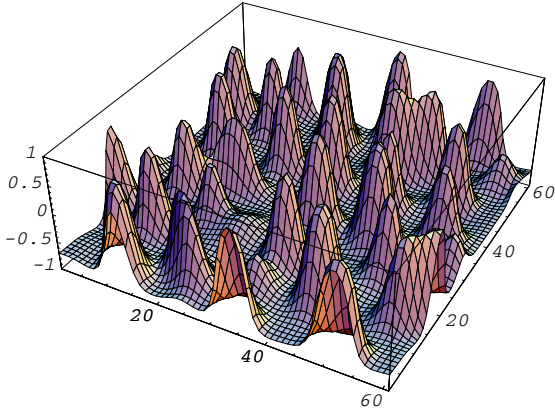}}
\caption[]{
  Order parameter in the transverse plane including expansion.
Expansion prevents droplets from merging.
}\end{figure}

These schematic calculations serve to illustrate the impact of droplet
formation on baryon fluctuations. To obtain more quantitative
predictions, one must use a more realistic form of the free energy
$f_0$. Our quadratic $f_0(\rho)$ strictly applies only near $T_c$ and
yields compressibilities that are equal at $\rho_h$ and $\rho_c$.
This result is unchanged if linear and cubic terms are added. In
contrast, lattice QCD calculations suggest that the compressibility
may jump across the transition \cite{Gavai}. The bag model equation of
state describes a first order transition and predicts a jump $
\Delta(\partial\rho/\partial\mu)\sim 2T_c^2$ for two light flavors, but is
not analytic in the two-phase region. We will discuss a more
sophisticated parameterization of $f_0(\rho)$ in future work.
\begin{figure} 
\epsfxsize=3.25in
\centerline{\epsffile{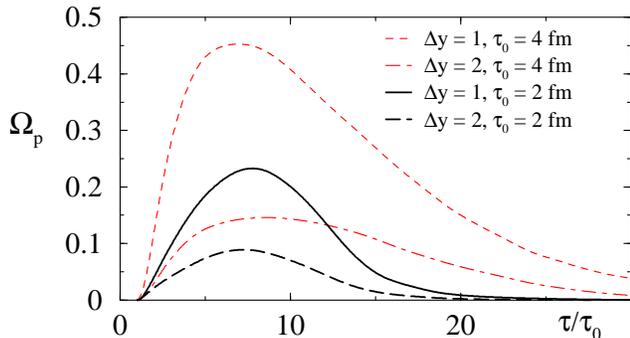}}
\caption[]{
  Enhanced variance vs. time for two rapidity windows, eq~(1).
}\end{figure}

Turning now to the interpretation of experiments, we emphasize that
the identification of phase-transition induced fluctuations requires a
systematic comparison of $\Omega_{p}$ in $pp$, $p$-nucleus ($pA$) and
nucleus-nucleus ($AA$) collisions. Benchmark RQMD and HIJING
simulations for central Au+Au collisions yield $\Omega_{p}\approx 0$
in the absence of a phase transition \cite{NuXu2}, in marked contrast
with fig.~4. However, we have been unable to find experimental results
on net-proton fluctuations in $pp$ or $pA$ collisions in the literature,
so it is not clear whether these benchmark estimates are reliable.  It
is therefore important that $AA$ experimenters study fluctuations of
identified baryons in $pp$ and $pA$ collisions. Measurements of
fluctuations of unidentified charged particles in hadronic collisions
\cite{Whitmore} and strange baryon production in $e^+e^-$ collisions
\cite{OPAL} hint of substantial proton-antiproton correlations in $pp$ 
collisions. If it turns out that light and heavy
ion fluctuations are similar, it may be necessary to correlate baryon
measurements with other signals to extract phase transition
information, as in \cite{Pruneau}.

Nevertheless, we stress that it is unlikely that super-poissonian
fluctuations in nucleon-nucleon ($NN$) collisions -- if present --
result in significant fluctuations in $AA$ interactions unless there is
a major source of coherence or collectivity. If we treat the $AA$
collision as a superposition of $NN$ subcollisions, then
$\Omega_{p}(AA) = \Omega_{p}(NN)/N(b)$,
where $N(b)$ is the number of participant nucleons. 
To obtain a rough upper bound on $\Omega_{p}(NN)$, we take the total
charge fluctuations measured to be $\sim 0.6$ in 200 GeV $pp$ collisions from
Whitmore's review
\cite{Whitmore}. For Au+Au collisions at $b < 10$~fm, we estimate
$\Omega_{p}(Au+Au) = \Omega_{p}(NN)/N(0) <
0.01$, where we use the wounded nucleon model to compute $N(b) \approx
59$ for $b = 10$~fm and 372 for $b = 0$. RQMD Au+Au simulations for
impact parameters fall below this bound \cite{NuXu2}.

We expect $\Omega_{p}$ to dramatically increase in heavy ion systems
compared to light ones. In central S+S we expect the $NN$ contribution
to $\Omega_{p}$ to be below 1\%, as implied by our wounded nucleon
model estimate. Since there is no evidence of a phase transition in
such light systems at AGS or SPS, the appearance of fluctuations at
the level of fig.~4 in Au+Au would be impressive.  But is there any
source of coherence or collectivity other than a phase transition?
Gluon junction effects \cite{Kharzeev} can lead to correlated baryon
production in $pp$, $pA$ and $AA$ collisions. This effect is only
partially included in RQMD \cite{Vance}. We are currently studying how
gluon junctions can effect $\Omega_{p}$ \cite{NuXu2}.

In summary, we have studied the phenomenological impact of baryon
density, a proposed order parameter of the putative first order QCD
phase transition at high baryon density \cite{Rajagopal}. We have
shown that phase separation in the nonequilibrium heavy ion system can
lead to large baryon fluctuations. These fluctuations are
super-poissonian and, consequently, can be extracted by measuring
protons alone. For (\ref{eq:phi4}) with $\langle \rho\rangle \propto
\tau^{-1}$, the system is unstable only for $\tau < 2.3\;\tau_0$. We
extend the calculations to much longer times to demonstrate that the
fluctuations in rapidity survive well past the freeze out time, of order
10--30 fm, in accord with \cite{Gavin}.

For sufficiently large $\tau_0$, final state fluctuations can be
substantial. However, we have seen that a more rapid expansion
corresponding to smaller $\tau_0$ leads to an ``inflation'' that
prevents the fluctuations from having a large impact on the final
state. If experiments find that the non-poissonian component of 
fluctuations is small, we must use information from flow signals to
ascertain the degree of this inflation.

We emphasize that these calculations include diffusion, which dampens
the fluctuations once the system becomes stable.  While diffusion is
the primary mechanism for dampening fluctuations at high density,
viscosity becomes more important at small net baryon density.  Several
key questions remain: At what energy do heavy ion collisions reach a
baryon density where the phase transition is strongly first order?  Is
there a residual modification of fluctuations due to the near
transition at zero baryon density?  To what extent does cooling,
convection, viscosity and collision-geometry alter $\Omega_p$ compared
to our estimates?  Finally, we note that our mixed-phase effect may be
compensated to some extent by the effect 
due to the difference between fluctuations in a plasma compared
to a hadron gas \cite{Koch}. Nevertheless, the strength of the signal in our
exploratory calculations invites further work.

We thank R. Bellwied, P. Braun Munzinger, K. Elder, M. Grant,
J. Kapusta, G. Kunde, P. Keyes, B. M\"uller, I. Mishustin, C. Pruneau,
S. Vance, S. Voloshin and N. Xu.  This work is supported in part by
the U.S. DOE grant DE-FG02-92ER40713.

\end{narrowtext}
\end{document}